\PassOptionsToPackage{prologue,dvipsnames}{xcolor}
\documentclass[sigconf, screen]{acmart}

\usepackage{booktabs}
\usepackage{amsmath,amsfonts}
\usepackage[outline]{contour}
\usepackage{fontawesome}
\usepackage{bm}
\usepackage{balance}
\usepackage{xspace}
\usepackage{float}
\usepackage{graphicx}
\usepackage{accents}
\usepackage{threeparttable}
\usepackage{diagbox}
\usepackage{textcomp}
\usepackage[skins]{tcolorbox}
\usepackage{enumitem}
\usepackage{graphicx}
\usepackage{multirow}
\usepackage{longtable}
\usepackage{url}
\usepackage[english]{babel}
\usepackage{ragged2e}
\usepackage{microtype}
\usepackage[caption=false]{subfig}
\usepackage{adjustbox}

\AtBeginDocument{
  \providecommand\BibTeX{{
    \normalfont B\kern-0.5em{\scshape i\kern-0.25em b}\kern-0.8em\TeX}}}
    
\DeclareUnicodeCharacter{2705}{\check}
\DeclareUnicodeCharacter{2014}{\cross}

\newcommand{\Code}[1]{\begin{small}\fontsize{8.5}{10}\selectfont\texttt{#1}\end{small}}

\newcommand{\tablesize}{\fontsize{8.0pt}{8.4pt}\selectfont}
\newcommand{\UpArrowLarge}{{\bgroup\contourlength{0.06em}\color{Green}{\contour{Green}{$\uparrow$}\egroup}}}
\newcommand{\UpArrowMedium}{{\bgroup\contourlength{0.03em}\color{Green}\contour{Green}{$\uparrow$}\egroup}}
\newcommand{\UpArrow}{{\color{Green}{$\uparrow$}}}
\newcommand{\DownArrowLarge}{{\bgroup\contourlength{0.06em}\color{Red}\contour{Red}{$\downarrow$}\egroup}}
\newcommand{\DownArrowMedium}{{\bgroup\contourlength{0.035em}\color{Red}\contour{Red}{$\downarrow$}\egroup}}
\newcommand{\DownArrow}{{\color{Red}{$\downarrow$}}}
\newcommand{\NoEffect}{\Code{-}}

\newlength{\fsize}

\tcbset{
  my box/.style={
    enhanced,
    colframe=#1!80,
    colback=#1!10,
    attach boxed title to top left={xshift=0.2cm, yshift=-0.2cm},
    boxed title style={
      colback=#1!80,
      outer arc=0pt,
      arc=0pt,
      top=0pt,
      bottom=0pt,
    },
  },
}
\newtcolorbox{result-rq}[1]{
  my box=black,
  title=#1,
  boxrule=1.2pt,top=6pt,bottom=3.5pt,left=6pt,right=6pt
}

\setcopyright{acmlicensed}
\acmPrice{15.00}
\acmDOI{10.1145/3611643.3616349}
\acmYear{2023}
\copyrightyear{2023}
\acmSubmissionID{fse23main-p1201-p}
\acmISBN{979-8-4007-0327-0/23/12}
\acmConference[ESEC/FSE '23]{Proceedings of the 31st ACM Joint European Software Engineering Conference and Symposium on the Foundations of Software Engineering}{December 3--9, 2023}{San Francisco, CA, USA}
\acmBooktitle{Proceedings of the 31st ACM Joint European Software Engineering Conference and Symposium on the Foundations of Software Engineering (ESEC/FSE '23), December 3--9, 2023, San Francisco, CA, USA}
\received{2023-02-02}
\received[accepted]{2023-07-27}

\begin{document}
\title{How Early Participation Determines Long-Term Sustained Activity in GitHub Projects?}

\author{Wenxin Xiao}
\affiliation{
  \institution{
    School of Computer Science, Peking University, and Key Laboratory of High Confidence
    Software Technologies, Ministry of Education}
  \city{Beijing}
  \country{China}
}
\email{wenxin.xiao@stu.pku.edu.cn}

\author{Hao He}
\affiliation{
  \institution{
    School of Computer Science, Peking University, and Key Laboratory of High Confidence
    Software Technologies, Ministry of Education}
  \city{Beijing}
  \country{China}
}
\email{heh@pku.edu.cn}

\author{Weiwei Xu}
\affiliation{
  \institution{
    School of Computer Science, Peking University, and Key Laboratory of High Confidence
    Software Technologies, Ministry of Education}
  \city{Beijing}
  \country{China}
}
\email{xuww@stu.pku.edu.cn}

\author{Yuxia Zhang}
\affiliation{
  \institution{
    School of Computer Science and Technology,
    Beijing Institute of Technology}
  \city{Beijing}
  \country{China}
}
\email{yuxiazh@bit.edu.cn}

\author{Minghui Zhou}
\authornote{Corresponding Author}
\affiliation{
  \institution{School of Computer Science, Peking University, and Key Laboratory of High Confidence
    Software Technologies, Ministry of Education}
  \city{Beijing}
  \country{China}
}
\email{zhmh@pku.edu.cn}

\begin{abstract}
    Although the open source model bears many advantages in software development, open source projects are always hard to sustain.
    Previous research on open source sustainability mainly focuses on projects that have already reached a certain level of maturity (e.g., with communities, releases, and downstream projects).
    However, limited attention is paid to the development of (sustainable) open source projects in their infancy, and we believe an understanding of early sustainability determinants is crucial for project initiators, incubators, newcomers, and users.    
    
    In this paper, we aim to explore the relationship between early participation factors and long-term project sustainability.
    We leverage a novel methodology combining the Blumberg model of performance and machine learning to predict the sustainability of 290,255 GitHub projects. Specificially, we train an XGBoost model based on early participation (first three months of activity) in 290,255 GitHub projects and we interpret the model using LIME.
    We quantitatively show that early participants have a positive effect on project's future sustained activity if they have prior experience in OSS project incubation and demonstrate concentrated focus and steady commitment. 
    Participation from non-code contributors and detailed contribution documentation also promote project's sustained activity. 
    Compared with individual projects, building a community that consists of more experienced core developers and more active peripheral developers is important for organizational projects. 
    This study provides unique insights into the incubation and recognition of sustainable open source projects, and our interpretable prediction approach can also offer guidance to open source project initiators and newcomers.
\end{abstract}




\maketitle

\section{Introduction}
\label{sec:intro}

Open Source Software (OSS) plays a fundamental role in both the computing world and our daily life.    
As the advantages of open source model are commonly recognized,
the number of OSS projects is rapidly growing.
According to the GitHub Octoverse Report from the last five years, about 32 million repositories were created in 2018, 44 million in 2019, 60 million in 2020, 61 million in 2021, and 86 million in 2022~\cite{GitHubOctoVerse}. However, a significant number of GitHub projects die in their first year~\cite{DBLP:conf/msr/AitIC22}. 
Some projects (e.g., course assignments) may have no intention of achieving long-term sustainability, but %previous studies show that 
even with these cases excluded, there are still a large amount of OSS projects suffering from long-term sustainability failures~\cite{DBLP:conf/sigsoft/CoelhoV17, DBLP:conf/sigsoft/ValievVH18, DBLP:conf/sigsoft/YinCXF21}. If we search over GitHub with the term ``is this dead''~\cite{Searchi97:online}, more than 329k issues will be returned, indicating an enormous interest in whether a project has been abandoned.
The large amount and the high failure rate of OSS projects pose formidable challenges to OSS project initiators, incubators, newcomers, and users.
Project initiators (i.e., developers who wish to launch and sustain their OSS projects) may face increased competition and decreased exposure to potential downstream users.
It will also be hard for incubators, newcomers, and users to locate promising OSS projects in their early stages to invest in, onboard early, or use.
Despite the great needs, guidelines for OSS project initiation come largely from the grey literature with little empirical evidence to support them~\cite{OSS-Start1, OSS-Start2, OSS-Start3, OSS-Start4}. 

A number of previous studies have investigated how various factors may affect long-term OSS project sustainability~\cite{chengalur2003survival, DBLP:journals/jais/Chengalur-SmithSD10, DBLP:journals/infsof/SamoladasAS10, DBLP:journals/jss/MidhaP12, wang2012survival, DBLP:journals/is/Ghapanchi15, DBLP:conf/sigsoft/ValievVH18, DBLP:conf/sigsoft/YinCXF21, yin2022code}.
However, they mainly focus on projects that have already reached a certain level of maturity (e.g., Apache Incubator projects~\cite{DBLP:conf/sigsoft/YinCXF21}) and investigate factors unavailable or unmeaningful for an OSS project in its early phase (e.g., developer and dependency networks~\cite{DBLP:conf/sigsoft/ValievVH18, DBLP:conf/sigsoft/YinCXF21,wang2012survival}, organizational ecology~\cite{DBLP:journals/jais/Chengalur-SmithSD10,chengalur2003survival}, number of downloads~\cite{DBLP:journals/jss/MidhaP12}). 
For factors available in the early phase, prior studies overwhelmingly focus on technical and process-related factors while overlooking participant-related factors~\cite{DBLP:journals/jss/MidhaP12, DBLP:journals/is/Ghapanchi15, yin2022code}.

In this paper, we argue that early participation plays an important role in determining long-term OSS project sustainability, analogical to how the initial effort of individuals determines the future sustainability of their work teams~\cite{houghton2003we, lindsley1995efficacy}.
To quantitatively confirm this, we ask the first research question (using sustained activity as the proxy for sustainability):

\begin{itemize}
    \item \textbf{RQ1:} To what extent can early participation factors predict an OSS project's long-term sustained activity?
\end{itemize}

To provide empirical evidence and interpretations for the predictability of early participation in sustained activity, we compare variables negatively/positively associated with sustained activity and ask the second research question:

\begin{itemize}
    \item \textbf{RQ2:} What early participation factors promote the long-term sustained activity of an OSS project?
\end{itemize}

To answer the \textbf{RQ}s, we sample 290,255 GitHub projects from GHTorrent~\cite{gousios2013ghtorent}, all of which demonstrate signals of continuous development in their first three months and rank top 5\% according to multiple popularity and activity metrics.
With reference to the Blumberg model~\cite{blumberg1982missing}, we design and extract a set of 54 variables from each project to measure the project's initial participation. % in the first three months of development.
%The variables are designed with reference to the Blumberg model~\cite{blumberg1982missing}. %a classic behavior model in management science.
We use these variables and ten other control variables to train an XGBoost model~\cite{DBLP:conf/kdd/ChenG16} that predicts whether the projects will have two years of sustained activity, achieving an AUC of up to 0.84.
To identify determinants of sustained activity, we further apply LIME~\cite{DBLP:conf/kdd/Ribeiro0G16} to generate local explanations for each prediction from the trained XGBoost model.
For each variable, we divide projects into two groups (where the variable correlates negatively/positively to sustained activity in the local explanations of LIME) and compare variable distributions between the two groups to understand the variable's contribution to sustained activity.

The above analysis reveals that early participation factors can indeed be used to predict project future sustained activity and several determinants play an important role.
We discover that the steady attention and commitment of core developers, participation of experienced developers (especially those with prior experience on projects with long-term sustained activity), and newcomer attractiveness have a positive effect on future sustained activity.
Organizational projects typically require more experienced core developers and more active peripheral developers than individual projects to achieve long-term sustainability. Based on our findings, we identify practices and strategies effective for incubating sustainable OSS projects, and provide signals that can be used to recognize long-term sustainable OSS projects. 
Our interpretable modeling approach can also serve as a starting point for the development of tools for OSS project initiation and selection.

\vspace{-1mm}
\section{Background and Related work}
\label{sec:related-work}

In the context of OSS development, the notion of sustainability is multifaceted~\cite{DBLP:conf/sigsoft/ValievVH18}, as is that of success~\cite{DBLP:conf/icis/CrowstonAH03}, and they share certain fostering factors (e.g., large team size can benefit both project sustainability and success~\cite{conlon2007examination}).
An OSS project's sustainability indicators may include:
1) artifact-related indicators~\cite{yin2022code} (e.g., following a modular and extensible architecture, high quality documentation); 
2) economic-related indicators~\cite{DBLP:journals/tse/ZhangZMJ21,DBLP:conf/sigsoft/ValievVH18,qiu2019going} (e.g., well funded, low cost of ownership, high added value); 
3) supply chain indicators~\cite{winters2020software, DBLP:conf/sigsoft/ValievVH18} (e.g., up-to-date dependencies, high number of downstream dependents);
4) development activity indicators~\cite{DBLP:conf/sigsoft/YinCXF21,2020FSE-Tan-First} (e.g., many commits, an adequately staffed team, influx of newcomers).
Among these indicators, sustained development activity is an especially strong signal as any piece of software will become progressively less useful if not maintained (i.e., Lehman's Law~\cite{lehman1980programs}).
Although development activity has been used as a proxy for both OSS success~\cite{DBLP:journals/jss/MidhaP12} and sustainability~\cite{DBLP:conf/sigsoft/ValievVH18}, Chengalur-Smith et al.~\cite{DBLP:journals/jais/Chengalur-SmithSD10} argue that sustainability requires activity to be maintained over the long run whereas success is measured at one particular time point.
In this paper, we use \textit{sustained activity} as the main proxy for studying OSS sustainability, following previous work~\cite{DBLP:conf/sigsoft/ValievVH18, DBLP:journals/is/Ghapanchi15,DBLP:conf/esem/AvelinoCVS19,DBLP:conf/esem/CoelhoVSS18}.

In recent years, the topic of OSS sustainability is gaining increasing public attention because a failed OSS project can generate a huge impact on our society.
This situation happens because many interdependent OSS projects form complex software ecosystems~\cite{DBLP:journals/ese/DecanMG19} and a large fraction of our modern digital infrastructure relies on such ecosystems~\cite{eghbal2016roads}.
Yet, even renowned OSS projects suffer from sustainability failures.
For example, the Log4Shell vulnerability in the popular Log4j 2 library affects over 17,000 Java libraries in Maven Central and countless Java applications~\cite{GoogleOn56:online}, but the library is painstakingly maintained by only four unpaid developers~\cite{6VolkanY51:online}.

Due to the high failure rate of OSS projects~\cite{DBLP:journals/jais/Chengalur-SmithSD10, DBLP:conf/msr/AitIC22}, OSS sustainability has been a long-lasting, multi-dimensional research topic with a large body of literature.
In the remainder of this section, we motivate our research based on two realms of prior work: the strategies for improving OSS sustainability and the determinants of OSS sustainability (at either project or individual level).

\textbf{Strategies for Improving OSS Sustainability.}
Sustainability failures can be caused by many factors~\cite{DBLP:conf/sigsoft/CoelhoV17, DBLP:conf/icse/CoelhoVSH18} (e.g., the departure of main contributors, usurped by competitors, etc.), and existing studies have focused on various strategies to overcome them.
For example, as a good influx of newcomers is often necessary to mitigate the risk of project abandonment~\cite{DBLP:conf/icsm/RehmanWKIM20,DBLP:conf/esem/AvelinoCVS19}, studies have investigated barriers to OSS newcomer onboarding~\cite{steinmacher2014barriers,shibuya2009understanding,DBLP:journals/infsof/SteinmacherSGR15,DBLP:conf/icse/Steinmacher0WG18} and explored strategies to help OSS newcomer onboarding (e.g., timely responses~\cite{2011HICSS-Carlos-Joining}, expert mentoring~\cite{tan2023enough,DBLP:journals/tse/ZhouM15}, university courses~\cite{he2023open}, and task recommendation~\cite{DBLP:conf/icsm/StanikMMFM18,2019Softw-Igor-Let,DBLP:conf/icse/XiaoHXTDZ22,2020FSE-Tan-First,xiao2023personalized}).
Apart from this line of work, studies have also shown how well-established foundations (e.g., the Apache Software Foundation~\cite{DBLP:conf/icse/YangFSA22}), business models~\cite{chang2007open}, and management~\cite{conlon2007examination} may promote long-term sustainability.

\begin{table}[t]
\definecolor{darkgreen}{rgb}{0.0, 0.55, 0.24}
\definecolor{darkred}{rgb}{0.55, 0.0, 0.0}
\footnotesize
\centering
\caption{Variables used in prior studies for modeling OSS project sustainability. Due to space constraints, we summarize their variables according to the high-level constructs used in their paper. For the used variables, we use \textcolor{darkgreen}{\faCheck} to mark variables available and meaningful during the early phase of an OSS project; we use \textcolor{red}{\faTimes} to mark variables otherwise.}
\vspace{-0.3cm}
\label{relcom}
\begin{tabular}{p{2cm}p{5.8cm}}
\toprule
    \textbf{Study} & \textbf{Variables used in the study}\\
\midrule
    \citet{chengalur2003survival} & \textcolor{darkgreen}{\faCheck}: size \newline \textcolor{red}{\faTimes}: age, reliability, niche \\
\midrule
    \citet{DBLP:journals/jais/Chengalur-SmithSD10} & \textcolor{darkgreen}{\faCheck}: development base size \newline \textcolor{red}{\faTimes}: project age, population niche size \\
\midrule
    \citet{DBLP:journals/infsof/SamoladasAS10} & \textcolor{darkgreen}{\faCheck}: \# of developers, project domain \\
\midrule 
    \citet{DBLP:journals/jss/MidhaP12} & \textcolor{darkgreen}{\faCheck}: license, \# of language translations, responsibility delegation, code complexity, code modularity \newline \textcolor{red}{\faTimes}: user base, developer base \\
\midrule 
    \citet{wang2012survival} & \textcolor{darkgreen}{\faCheck}: license, external network size, external network quality, targeted users, user/developer participation \newline \textcolor{red}{\faTimes}: internal network size, service quality\\
\midrule
    \citet{DBLP:journals/is/Ghapanchi15} & \textcolor{darkgreen}{\faCheck}: defect removal, functionality enhancement \newline \textcolor{red}{\faTimes}: release frequency \\
\midrule
    \citet{DBLP:conf/sigsoft/ValievVH18} & \textcolor{darkgreen}{\faCheck}: organization involvement,  \# of upstream dependencies \newline \textcolor{red}{\faTimes}: \# of downstream dependencies, indirect connectivity through transitive dependencies, backporting \\
\midrule
    \citet{DBLP:conf/sigsoft/YinCXF21} & \textcolor{darkgreen}{\faCheck}: \# of developers, \# of commits, \# of emails, \# of files, \# of interruptions, \% of top contributions \newline \textcolor{red}{\faTimes}: (internal) social network, technical network\\
\midrule
    \citet{yin2022code} & \textcolor{darkgreen}{\faCheck}: code (e.g., lines of code, \# of directories), process (e.g., \# of major/minor/new contributors, \# of files added/deleted), quality (e.g., tests, code complexity, function size) \\
\midrule  
    Our Study & \textcolor{darkgreen}{\faCheck}: participant willingness (core/peripheral/non-code), capacity (core/peripheral/non-code), and opportunity \\
\bottomrule
\end{tabular}
\vspace{-4mm}
\end{table}

\textbf{Determinants of OSS Sustainability.} 
To help stakeholders take timely and proactive action against sustainability risks, a number of studies have modeled OSS project sustainability in different settings and identified sustainability determinants from different perspectives~\cite{chengalur2003survival, DBLP:journals/jais/Chengalur-SmithSD10, DBLP:journals/infsof/SamoladasAS10, DBLP:journals/jss/MidhaP12, wang2012survival, DBLP:journals/is/Ghapanchi15, DBLP:conf/sigsoft/ValievVH18, DBLP:conf/sigsoft/YinCXF21, yin2022code}. The main variables of interest in prior studies are summarized in Table~\ref{relcom}. We can observe that many variables in prior studies are unavailable in the early phases of an OSS project (e.g., population niche~\cite{DBLP:journals/jais/Chengalur-SmithSD10}, user and developer base~\cite{DBLP:journals/jss/MidhaP12}, dependency network metrics~\cite{DBLP:conf/sigsoft/ValievVH18}, and social-technical network metrics~\cite{DBLP:conf/sigsoft/YinCXF21}).
Consequently, findings related to these variables are only applicable to projects with established communities (i.e., a socio-technical network) or reached milestones (e.g., releases, artifacts, downstream projects).
Even for the variables available during the early phase, prior studies extensively focus on technical and process-related variables (e.g., code complexity~\cite{DBLP:journals/jss/MidhaP12, yin2022code}, license~\cite{DBLP:journals/jss/MidhaP12, wang2012survival}, features and bug fixes~\cite{DBLP:journals/is/Ghapanchi15, yin2022code}) while overlooking participant-related factors.
The only exception is \citet{wang2012survival}, which investigated the amount of participation and the quality/size of the external network (as a measure of available resources and social capital) at a coarse granularity.
On the other hand, participants, especially early participants, can play a key role in sustaining OSS projects, as evidenced by previous studies on OSS participant motivation~\cite{10.5555/776816.776867,Krogh2012Carrots,DBLP:conf/icse/GerosaWTLRTSS21} and the modeling of individual sustainability~\cite{DBLP:journals/jmis/FangN09, 2012ICSE-Zhou-What, DBLP:journals/tse/ZhouM15, DBLP:conf/icgse/0008RS17, qiu2019going, DBLP:journals/tse/BaoXLM21} (i.e., what drives an individual's sustained participation in OSS projects).
For example, \citet{2012ICSE-Zhou-What, DBLP:journals/tse/ZhouM15} empirically show how the initial behaviors and experiences of OSS participants correlate to their sustained participation.
%To the best of our knowledge, no previous study has investigated in depth the correlation between initial participant factors and long-term OSS project sustainability.
However, we are unaware of any prior study carrying out an in-depth investigation between early participant factors and long-term \textit{project} sustainability, leaving an important knowledge gap relevant for many stakeholders during OSS project incubation (e.g., initiators, incubators, newcomers, and users).\footnote{
Note that project sustainability and individual sustainability are two different outcomes for which project sustainability is the pre-condition of individual sustainability. Studies of individual sustainability are typically conducted in long-term sustainable projects, whose findings may not be applicable to OSS project incubation.
}

In this paper, we fill this gap by proposing an interpretable modeling approach that predicts future sustained activity using early participant factors.
Our study takes a novel perspective showing that early participation factors can effectively predict long-term project sustainability and reveal interesting correlations. 
We will discuss in Section~\ref{secimplication} how filling this knowledge gap strengthens previous findings and yields new implications for practitioners.

\section{Methodology}
\label{sec:approach}

\begin{figure}
    \centering
    \includegraphics[width=\linewidth]{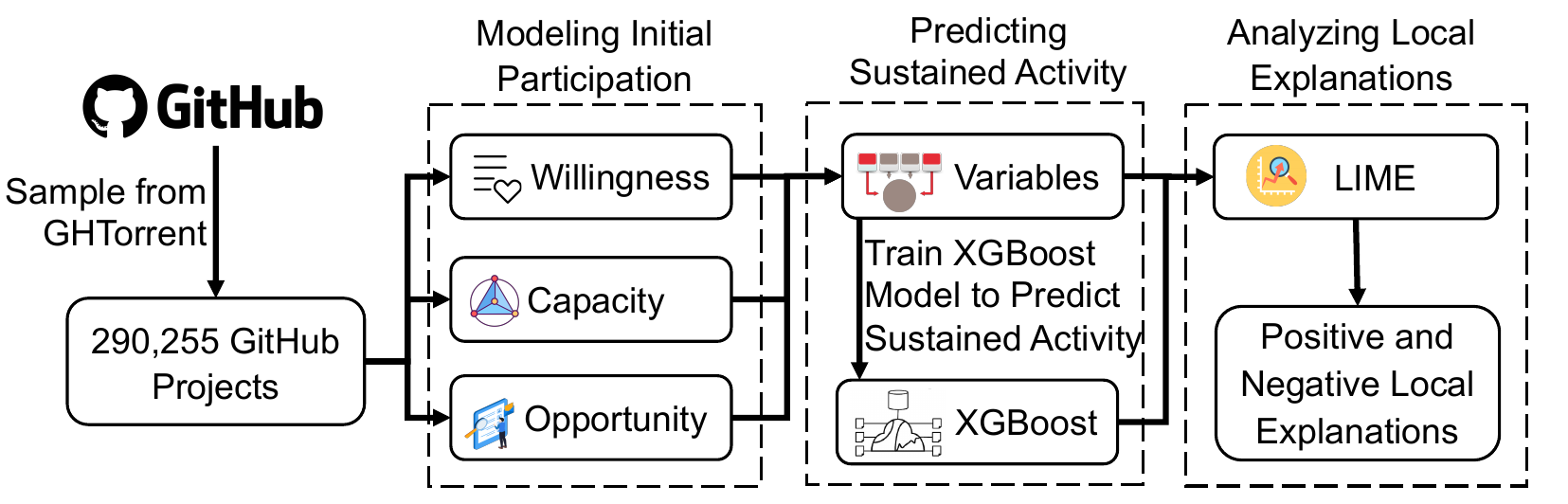}
\vspace{-6mm}
    \caption{An overview of the study methodology.}
\vspace{-4mm}
    \label{fig:overview}
\end{figure}

In this Section, we first %formulate research questions (\textbf{RQ}s) and 
describe how a sample of GitHub projects is collected from GHTorrent~\cite{gousios2013ghtorent} to conduct this study.
Then, we design variables to measure early participation w.r.t. the Blumberg model of performance~\cite{blumberg1982missing}.
Next, we introduce how XGBoost~\cite{DBLP:conf/kdd/ChenG16} is employed to predict two-year sustained activity based on the variables.
Finally, we explain how the LIME~\cite{DBLP:conf/kdd/Ribeiro0G16} is applied to explain the correlation between variables and the prediction (i.e., two-year sustained activity) in the trained XGBoost model.
An overview of our study methodology is available in Figure~\ref{fig:overview}.

\subsection{Project Selection}\label{sec:prosl}

In this study, we focus on GitHub projects and collect project data from the latest GHTorrent dump (version 2021-03-06)~\cite{gousios2013ghtorent}. 
GHTorrent provides event timeline data allowing us to reproduce historical project states and collect project variables at any time point, which have also been used in 
previous software engineering studies to retrieve the historical data of GitHub projects~\cite{DBLP:conf/msr/KalliamvakouGBSGD14,DBLP:conf/icse/XiaoHXTDZ22}.

There are 189,467,746 projects (i.e., repositories) recorded in GHTorrent.
Similar to the dataset filtering process adopted by \citet{DBLP:journals/ese/MunaiahKCN17}, we first remove deleted projects and forks of other projects.
Forks are removed because this study focuses on the incubation of a \textit{new} project at its initial stage (note that contributions successfully merged from forks will be considered).
%GitHub became popular since 2012~\cite{DBLP:conf/sigsoft/ValievVH18}. 
Then, we filter out projects created before 2012 as GitHub only became popular after 2012~\cite{DBLP:conf/sigsoft/ValievVH18}; we also filter out projects created after March 2019 as their two-year sustained activity status cannot be observed from our dataset. 
To filter out projects with no intention of sustaining long-term development (e.g., toy projects, classwork projects, paper repositories, and projects for competitions), we only retain projects with a certain level of activity, popularity, and development time spans. 
Specifically, we compute the 95th percentile of all projects in GHTorrent for the following popularity and activity metrics: number of commits, pull requests, issues, forks, and stars; we only retain projects whose metrics are no less than the 95th percentile, i.e., with no less than 57 commits, 4 PRs, 1 issue, 1 fork, and 2 stars, before the time the GHTorrent dump was collected.
Then, we further retain projects with a time span of at least three months between the first and last commits which could indicate some intention of sustained development. %(only $\sim$20\% of the projects were developed for more than three months).
After the above filtering steps, the final dataset contains 290,255 projects.
%Intuitively, the sampled projects are less likely to be toy projects and tend to have long-term development intentions.
Although the above sampling heuristics have limitations (to be discussed in Section~\ref{sec:limitations}), the resulting dataset contains mostly engineered GitHub projects~\cite{DBLP:journals/ese/MunaiahKCN17} with only a few cases of noise (e.g., classwork).\footnote{
To validate dataset quality, we sample 384 projects (confidence level: 95\%, margin of error: 5\%) 
and check whether they are engineered software projects following the labeling process in \citet{DBLP:journals/ese/MunaiahKCN17}.
Among them, 19 have been deleted at the time of writing, six store documents, and three are for classwork.
In other words, 92.7$\pm$5\% of the projects in our dataset are engineered software projects (confidence level = 95\%).
} 
%Our aim is to assist such projects that have long-term development intentions but face failure.

\begin{comment}
\begin{figure}[b]
\centering
\vspace{-4mm}
\includegraphics[width=0.8\linewidth]{pro_eachyear.png}
\vspace{-1mm}
\caption{Number of projects created each year in our dataset.}
\label{fig:staend}
\end{figure}
\end{comment}

We observe in our dataset that it is common for seemingly abandoned projects to restart development after a period of time.
This observation is consistent with the results of Coelho and Valente~\cite{DBLP:conf/sigsoft/CoelhoV17}, that projects may switch ownership or accept new core developers to get rid of the unmaintained status.
Thus, it is inadequate to measure the sustained activity of a project with only the first and last commits~\cite{DBLP:conf/sigsoft/ValievVH18}. 
In this study, we consider a project as having \textit{$t$-year sustained activity} if 1) they have commit activity for more than $t$ year(s), and 2) the median number of commits per month is at least $k$. $k$ is an adjustable variable, and a higher value of $k$ indicates a stricter definition of projects with sustained activity.

\begin{table}[b]
\small
\centering
\vspace{-0.4cm}
\caption{The ratio/number of projects started each year with at least two years of sustained activity (``Two-Year'' column) / still active at dataset truncation (``Still Active'' column).}
\vspace{-3mm}
\label{tab:age}
\begin{tabular}{ccccccc}
\toprule
& &\multicolumn{2}{c}{Ratio of Projects}
&\multicolumn{2}{c}{Number of Projects}\\
\midrule
Start & Sum & Two-Year & Still Active & Two-Year & Still Active\\
\midrule
2012 & 17,056 & 0.46 & 0.10 & 7,817 & 1,779 \\
2013 & 27,409 & 0.48 & 0.11 & 13,110 & 2,996\\
2014 & 36,226 & 0.45 & 0.12 & 16,359 & 4,404 \\
2015 & 46,954 & 0.43 & 0.14 & 20,121 & 6,458  \\
2016 & 52,111 & 0.43 & 0.17 & 22,159 & 9,086  \\
2017 & 53,863 & 0.40 & 0.23 & 21,278 & 12,399  \\
2018 & 49,108 & 0.39 & 0.33 & 19,252 & 16,235 \\
2019 & 7,528 & 0.33 & 0.33 & 2,471 & 2,497  \\
\midrule
\textsc{Sum} & 290,255 & 0.42 & 0.19 & 122,567 & 55,854  \\
\bottomrule
\end{tabular}
\end{table}

We regard the first commit time of a repository as its start time. Table~\ref{tab:age} shows the survival status of projects started each year from 2012 to March 2019. The number of repositories started each year that meet our filtering rules peaked in 2017. 
%Despite the large number of projects that were created in recent years, the percentage of projects that subsequently reach two years of sustainability is declining.
For $t=2$ and $k=1$, we present the ratio and number of projects with sustained activity (i.e., the \textit{Two-Year} column), and projects that are still active within two years before data truncation (i.e., the \textit{Still Active} column), in Table~\ref{tab:age}.
We observe that the ratio of two-year sustainable projects gradually decreases over years, despite the growing number of new GitHub projects per year~\cite{GitHubOctoVerse}.
The ratio of still active projects in the last three years is relatively high and fluctuated around 13\% before 2017.
Besides, the top 10 most widely used programming languages on GitHub~\cite{GitHubOctoVerse} are also the most frequently used languages among the selected projects, comprising the top 13 languages. The number of commits for these projects ranged from 57 to 3,545,589, with a median value of 190 and a mean value of 571.

\vspace{-0.5mm}
\subsection{Measuring Early Participation}
\label{sec:feature}
\vspace{-0.25mm}

In GHTorrent, about 1/4, 1/5, and 1/6 of the projects on GitHub have commit activity over spans of one, three, and five month(s).
In this study, we measure early participation using development activities from the first $m=1,3,5$ month(s), respectively.
%The choice of three months is rather arbitrary with the intuition that three months would be sufficient for most projects to kickstart themselves, and only 20\% of GitHub projects are developed for the first three months (Sec.~\ref{sec:prosl}).
We borrow the model of work performance proposed by Blumberg and Pringle~\cite{blumberg1982missing} to identify different dimensions of variables that affect the performance of participants from three different groups: core developers, peripheral developers, and non-code contributors.

\textbf{The Blumberg Model of Performance.}
As a classic behavior model in management science, the Blumberg model of performance proposes that three interacting factors influence the performance and/or behavior of an individual: willingness, capacity, and opportunity. Among them, willingness refers to psychological antecedents including personality, attitude, motivational effect, etc.; capacity consists of physiological and cognitive determinants such as experience and ability; opportunity includes social or technological factors that constrain or enable people's work behavior in an organizational setting. 
This three-dimensional framework has been used in software engineering studies to predict the likelihood of a developer becoming a long-term contributor in OSS projects, explore why newcomers abandon OSS projects, and study developer role evolution in OSS ecosystems~\cite{2012ICSE-Zhou-What,DBLP:conf/icse/SteinmacherWCG13,DBLP:journals/jcst/ChengLLZL17}.

\textbf{Core Developers, Peripheral Developers, and Non-Code Contributors}.
%Because repository-related developers with different roles,
Participants in an OSS project may serve different roles (e.g., code contributors, non-code contributors, users, potential newcomers) and the tasks they perform vary greatly~\cite{DBLP:conf/icse/JensenS07,2012ICSE-Zhou-What,DBLP:journals/ese/IzquierdoC22}.
In previous work, code contributors in OSS projects are classified into core and peripheral roles~\cite{DBLP:conf/hicss/CrowstonWLH06,DBLP:journals/tosem/MockusFH02}, in which core developers are most productive and play a leading role in project development and maintenance~\cite{DBLP:conf/iwpse/YamashitaMKHU15,DBLP:journals/tosem/MockusFH02}.
However, other contributors also have their roles to take, i.e., \textit{given a large enough beta-tester and co-developer base, almost every problem will be characterized quickly and the fix obvious to someone}~\cite{DBLP:journals/firstmonday/Raymond98}. 
In this study, we follow the most common operationalization and distinguish among three types of participants based on their commits to the project.
Specifically, we consider the participants who made the most commits and whose commit counts add up to over 80\% of the total as \textit{core developers}; the remaining participants with at least one commit are regarded as \textit{peripheral developers}~\cite{DBLP:conf/icse/JoblinAHM17,DBLP:journals/tosem/MockusFH02}; participants who only commented on issues, pull requests (PRs) or reported issues are considered as \textit{non-code contributors}, who may give ideas for project future evolution, influence development, and enrich discussions~\cite{DBLP:journals/ese/IzquierdoC22}.
For each project, we compute participant-level variables as the average of all participants in each group, and we expect different groups to have different impacts on long-term sustained activity.

In the remainder of this Section, we introduce how the variables for modeling early participation are defined w.r.t. the three dimensions of willingness, capacity, and opportunity. Regarding willingness, we focus on variables that represent the strength and trend of willingness. As for capability, we consider both general OSS capability and project-specific capability. For opportunity, we focus on opportunity for newcomers to participate in the project. 
We also consider indirectly related control variables.
Table~\ref{tab:stat1} summarizes all variables and their definitions.

\textbf{\textit{Willingness of Participants.}}
Early investment indicates intrinsic motivations to participate, which drives developers to sustain contributions in the future~\cite{DBLP:journals/mansci/RobertsHS06}. Following previous studies~\cite{2012ICSE-Zhou-What,DBLP:conf/ifip5-5/AsriKBJ17,konchady2016querying} that use effort spent on a project as the main proxy for developers' willingness to long-term/actively contribute, we use the effort in terms of \textbf{cumulative amount} and \textbf{stability} to measure the strength and trend of developers' willingness.
We suspect that developers whose attention is spread across multiple projects may not be as inclined to focus their energy on a new project. Thus we also use developers' \textbf{level of concentration} to laterally measure their willingness to incubate projects.

%We measure the willingness of participants from three sub-dimensions: the effort spent, the general interest shown, and the pattern of contribution.
\vspace{-1mm}
\begin{itemize}[leftmargin=20pt]
\item[\scriptsize{WP}\small{1}]
\textbf{Cumulative Effort Amount:} As different types of contributions (e.g., commits, issues, and comments) require varying amounts of effort~\cite{2012ICSE-Zhou-What,konchady2016querying}, we measure developers' efforts in terms of the average number of different types of contributions made by each group of participants to the project, namely, number of commits %(\Code{\#cmt\_\{c|p\}})
, PRs% (\Code{\#pr\_\{c|p\}})
, issues% (\Code{\#issue\_\{c|p|n\}})
, issue comments% (\Code{\#iss\_comment\_\{c|p|n\}})
%, number of PR comments% (\Code{\#pr\_comment\_\{c|p|n\}})
, commit comments% (\Code{\#cmt\_comment\_\{c|p|n\}})
, and issue events% (\Code{\#iss\_event\_\{c|p|n\}})
. %such as ``mentioned'' and ``labeled''.

\item[\scriptsize{WP}\small{2}]
\textbf{Stability of Effort:} We assume that declining commitment may mean a weakened willingness to continue development and a steady stream of contributions implies participants' willingness to sustain project development in the long term.
Thus, we measure the contribution ``steadiness'' using the following variables: 
the number of days with commits  %(\Code{\#cmt\_actday})
and the number of commits in the first/second half of observation period%first/last 45 days (\Code{\#cmt\_front}/\Code{\#cmt\_end})
;
the median %(\Code{\#cmt\_median}) 
and standard deviation %(\Code{cmt\_day\_std}) 
of the commit numbers for each day; 
and the standard deviation % (\Code{cmt\_dev\_std}) 
of the number of commits made by all code contributors. %In addition, we calculate median PR merge time (\Code{\#merge\_t}) for each project. Note that the features in this module are only measured for code contributors.\\

\item[\scriptsize{WP}\small{3}]
\textbf{Level of Concentration:} 
Developers may follow each other and star a repository for multiple reasons~\cite{DBLP:conf/icse/SingerFS14,DBLP:journals/jss/BorgesV18} (e.g., to bookmark or to show appreciation~\cite{DBLP:journals/jss/BorgesV18}). 
If a developer follows many projects or developers, it may indicate their broad curiosity, follow-ups to new technology, or divergent attention. Such a developer may be more interested in and willing to participate in multiple mature projects, rather than dedicating as much time and effort as possible to a startup project.
Therefore, we use the number of developers followed by participants and the number of projects they starred to measure their general level of concentration on GitHub.
\end{itemize}
\vspace{-1mm}

\textbf{\textit{Capacity of Participants.}}
Historical experience is a commonly used proxy to measure developer expertise in previous studies (e.g., code reviewer and crowdsourcing developer recommendation~\cite{DBLP:conf/icse/WangY0HWW20,DBLP:conf/icse/RahmanRC16}). Therefore, we measure how much and how well developers contribute on GitHub by their \textbf{general OSS experience} including the cumulative volume of contributions and experience of successfully incubating projects with long-term sustained activity. 
In addition, specific capabilities required by the project can be directly measured through code-related variables. Stănciulescu et al.~\cite{yin2022code} have identified code-related variables that may be associated with project sustainability, which will be discussed in detail in Section~\ref{alternatives}.
Meanwhile, developers are usually followed because they are recognized for their historical work, so we use \textbf{popularity} to portray developers' capability.

%We measure the capacity of participants from two sub-dimensions: their general OSS experience and their popularity among other GitHub users.
\vspace{-1mm}
\begin{itemize}[leftmargin=20pt]
\item[\scriptsize{CP}\small{1}]
\textbf{General OSS Experience:} 
Developers with rich experience in OSS development and the successful incubation of sustainable projects may be more capable of fostering a project with long-term sustained activity. 
As previous work shows that experience can be measured in terms of cumulative task performance~\cite{DBLP:journals/orgsci/ArgoteM11}, we extract the following variables to characterize the historical OSS experience of project participants: the number of commits, % (\Code{\#cmt\_all\_\{c|p|n\}})
 issues% (\Code{\#issue\_all\_\{c|p|n\}})
, and PRs %(\Code{\#pr\_all\_\{c|p|n\}}) 
on GitHub, number of owned projects% (\Code{\#pro})
, and number of owned projects with more than one or two years of sustained activity% (\Code{\#pro\_oneyear}, \Code{\#pro\_twoyear})
.

\item[\scriptsize{CP}\small{2}]
\textbf{Popularity:}
Developers are often followed because other developers consider their historical work interesting or useful.
%Therefore, we assume that developers' popularity may signal their ability to develop useful or interesting projects, and use the number of followers %(\Code{\#follower\_\{c|p|n\}}) 
%to measure their popularity. 
Therefore, we use the number of followers to measure developers' ability to develop useful or interesting projects.
\end{itemize}
\vspace{-1mm}

\textbf{\textit{Opportunity of Contribution.}}
A major reason for OSS sustainability failures is the lack of time/interest from core developers and the absence of new developers~\cite{DBLP:conf/sigsoft/CoelhoV17}, and an effective way to avoid such failure is to provide opportunities for newcomers to contribute and support their onboarding.
Specifically, developers may have more chances to contribute if there are more available tasks from open issues. 
We are also interested in the impact of using \emph{good first issue} labels, an increasingly popular label intended for marking newcomer-friendly tasks~\cite{2020FSE-Tan-First}. In addition, projects usually provide documents such as README.md and CONTRIBUTING.md to help newcomers quickly understand project background and contribution workflow. Therefore, we use the number and ratio of open issues, the number of \emph{good first issues}, the number of lines of README and CONTRIBUTING documentation to measure the contribution opportunity of potential newcomers. 
It should be noted that open issues only represent potential opportunities for newcomers, and they may not be suitable for all newcomers, because some newcomers may lack the expertise to resolve the issues.

\textbf{\textit{Control Variables.}}
Some variables are related to inherent properties of developers or projects (e.g., development team size, project popularity, project owner account type); they do not characterize early participation but have been shown to be associated with project sustainability~\cite{DBLP:journals/jais/Chengalur-SmithSD10,DBLP:conf/sigsoft/ValievVH18}. 
Thus, we include the following control variables in our study: the number of members, the number of forks, project owner account type, the number of organizations participants belong to, and the ratio of participants showing affiliated companies/institutions on their overview page. 

\vspace{-1mm}
\subsection{Training an XGBoost Model}
\label{sec:intermodel}

Given the collected variables, our next task is to train a model to effectively predict the project's $t$-year sustained activity.
The problem is basically a binary classification task.
We define project status (\Code{status}) as the prediction outcome: if a project has sustained activity for more than $t$ year(s), we set $\Code{status=1}$, and otherwise $\Code{status=0}$.
In this study, we use XGBoost~\cite{DBLP:conf/kdd/ChenG16}, a library for gradient-boosted regression tree learning with clever optimization tricks and efficient implementations, for learning and prediction.
We choose to use XGBoost because: 1) it is highly flexible, supporting automated feature selection and the handling of both numerical and categorical variables;
2) it has exhibited competitive predictive performance in diverse software engineering prediction tasks~\cite{DBLP:journals/ase/SantosFVVZ20, DBLP:conf/iwpc/Yazdaninia0S21, DBLP:conf/icse/XiaoHXTDZ22, GFI-Bot}; and 3) the learned model also outperforms two alternative machine learning models (i.e., logistic regression and random forest) in our dataset in terms of AUC (as we will show in Section~\ref{ss:predict}).
With current performance, we believe the model is able to learn some generalizable patterns worthy of further analysis and interpretation.

\vspace{-1mm}
\subsection{The Interpretable LIME Approach}
\label{lime}

To investigate how variables contribute to the prediction results of the XGBoost model, we further interpret the results using the Local Interpretable Model-agnostic Explanations (LIME), which is designed to explain the prediction results yielded by any pre-trained black-box machine learning model~\cite{DBLP:conf/kdd/Ribeiro0G16}. 
LIME has two advantages: one is that it is model-agnostic, i.e., applies to arbitrary linear or nonlinear classifiers; the other is that it allows us to locally analyze the variable contributions of individual samples by local fitting. 

For complex nonlinear systems that are ubiquitous in reality, the coefficient of a variable is determined by its position in the system~\cite{enns2011s}; that is, when the values of other variables are different, even if the value of this variable does not change, it may have different coefficients. 
Thus, a variable with the same value may have opposite effects on different projects.
For example, as shown in Figure~\ref{vardistibution}, the standard deviation of commit number per code contributor $\Code{cmt\_dev\_std}=4$ is more likely to have a negative effect on sustained activity when $\Code{type}=0$ (the project is owned by an organization), and is positive for sustained activity when $\Code{type}=1$ (details are in Sec.~\ref{sec:rq2}). 
We use the local explanations generated by LIME from the XGBoost model to analyze the contribution of variables to sustained activity in different project contexts.

Given a trained model and a data point, LIME generates neighborhood data points for that sample and learns a linear model locally on these data points, so that the contribution of each variable to the prediction result is locally interpretable within the linear model. 
The coefficient of one variable for a data point can be either positive or negative, indicating that the variable contributes positively or negatively to $t$-year sustained activity in this data point.
For each variable $v$, to analyze the overall contribution of $v$ to the prediction results on the whole dataset, we divide projects into two groups: the \textit{negative group} where LIME returns negative coefficients for $v$ and the \textit{positive group} where LIME returns positive coefficients for $v$.
The overall value distributions of $v$ in the two groups are compared using the Mann–Whitney U test~\cite{mann1947test} 
with Bonferroni correction~\cite{dunn1961multiple}.
We also compute the effect size ($|r|$) and consider the effect size as large when $|r| \ge 0.5$, medium when $|r| \ge 0.3$, small when $|r| \ge 0.1$, and negligible otherwise, following common recommendations in applied statistics~\cite{fritz2012effect}.
A variable has a large/medium/small positive effect on sustained activity if the median of $v$'s positive group is larger than the median of $v$'s negative group and its effect size is greater than 0.5/0.3/0.1, and vice versa.

\vspace{-1mm}
\section{Results}
\label{sec:res}
\vspace{-0.5mm}

\subsection{RQ1: Predictive Performance}
\label{ss:predict}
\vspace{-0.5mm}

We evaluate the prediction performance of our model under different $m, t$ and $k$. Here, the data of projects are collected at their first $m$ month(s) and $t$-year sustained activity is defined with a hyperparameter $k$. 
We also compare our model with a number of baselines to demonstrate the competitiveness of XGBoost on this problem and investigate the contribution of each variable dimension.

\vspace{-2mm}
\subsubsection{Evaluation Metrics}
In our study, we consider the Area Under the ROC Curve (AUC) as the main evaluation metric because it can holistically measure the goodness of fit and prediction effectiveness at all classification thresholds.
Intuitively, AUC denotes the probability that a random positive sample ranks higher than a random negative sample, and it is generally more appropriate to evaluate classifiers with imbalanced data~\cite{DBLP:books/sp/FernandezGGPKH18}. 
We also provide precision and recall at the classification threshold $0.5$, as a reference. 

\vspace{-2mm}
\subsubsection{Experimental Performance under Different Scenarios}
\label{diffset}
The lower quartile, median, and upper quartile of monthly commit counts for projects in our dataset with more than two years of sustained activity are 1, 2, and 6, respectively. 
Table~\ref{tab:k} presents the performance of our approach in predicting $t=1$ and $2$ year(s) of sustained activity when the median number of commits per month in the first $t$ year(s) is at least $k=1,2$ and $6$ with projects' data collected from the first $m=1,3,5$ month(s) (about 1/4, 1/5, and 1/6 of the projects in GHTorrent database are developed longer than them, respectively), reported from 10-fold cross-validation~\cite{stone1974cross}.

\begin{table}[h!t]
\footnotesize
\centering
%\vspace{-0.3cm}
\caption{AUC/Precision/Recall under different parameters.}
\vspace{-0.3cm}
\label{tab:k}
\begin{tabular}{c c c c c c c}
\toprule
&$m$&$k=1$&$k=2$&$k=6$\\
\midrule
& 1 & 0.65 / 0.60 / 0.59 & 0.66 / 0.58 / 0.39 & 0.72 / 0.57 / 0.18 \\
$t = 1$ & 3 & 0.73 / 0.65 / 0.69 & 0.74 / 0.63 / 0.56 & 0.81 / 0.62 / 0.38 \\
& 5 & 0.77 / 0.68 / 0.75 & 0.80 / 0.67 / 0.65 & 0.86 / 0.66 / 0.53 \\
\midrule
& 1 & 0.66 / 0.59 / 0.41 & 0.68 / 0.58 / 0.24 & 0.73 / 0.56 / 0.13 \\
$t = 2$ & 3 & 0.72 / 0.62 / 0.52 & 0.74 / 0.61 / 0.38 & 0.80 / 0.60 / 0.26 \\
& 5 & 0.76 / 0.65 / 0.59 & 0.78 / 0.64 / 0.47 & 0.84 / 0.63 / 0.38 \\
\bottomrule
\end{tabular}
\vspace{-0.3cm}
\end{table}

Obviously, the model can predict better with data collected over longer time horizons (higher $m$). When $m=3$, moderately accurate (0.7$<$AUC$\leqslant$0.9)~\cite{greiner2000principles} can already be achieved for all $k$s and $t$s. 
Though $t=2$ is a longer period for ``sustained development'', the model's performance in predicting two-year sustainability is slightly lower than that of one year, which is also expected as the model needs to predict a more distant future.
At larger $k$, AUC becomes better, maybe because active projects are more easily identified, but precision and recall become worse.
Due to space constraints, we only report the results for $m=3$, $t=2$, and $k=1$ in the following experiments. 
The results are stable across other parameter settings and are provided in the replication package.

\vspace{-2mm}
\subsubsection{Comparison with Alternatives}\label{alternatives}
To demonstrate the contribution of our variables and XGBoost to the final performance, we compare our approach with several alternative baselines. % of our approach. 
 
\vspace{-0.5mm}
\begin{itemize}[leftmargin=10pt]
    \item \textbf{Alternative ML Models:}
    \textit{Logistic Regression} and \textit{Random Forest} are popular models with widespread use in software engineering~\cite{DBLP:conf/icsm/StanikMMFM18,DBLP:conf/sigsoft/ValievVH18}. Logistic regression is especially popular for empirical analysis~\cite{DBLP:conf/sigsoft/ValievVH18,2012ICSE-Zhou-What}. We employ both models from the \Code{scikit-learn} Python package using all our variables as two baselines to showcase the competitiveness of our approach and substantiate its adoption over traditional data analysis approaches.
    \item \textbf{Isolated Feature Dimension:} 
    Among the variables we propose, 
    \Code{\#cmt\_\{c|p\}}, \Code{\#issue\_\{c|p|n\}}, \Code{\#cmt\_actday}, 
    \Code{type}, and \Code{\#member} are also studied by previous work shown in Table~\ref{relcom}.
    To demonstrate the effect of these variables, we show the performance of two alternative XGBoost models with only these common variables (i.e., \textit{Common}) and with all other variables (i.e., \textit{Other}), respectively.  
    As introduced in Section~\ref{sec:feature}, the variables are collected from seven sub-dimensions, i.e., \textit{Cumulative Effort Amount, Stability of Effort, Level of Concentration, General OSS Experience, Popularity, Opportunity of Contribution}, and \textit{Control Variables}.
    To examine the predictability of variables from each dimension on projects' sustained activity, we show the performance of seven alternative XGBoost models, each of them using only variables from one of the seven sub-dimensions. 
    \item \textbf{Previous Study:} We compare our approach with \textit{Stănciulescu et al.}~\cite{yin2022code}, who proposed using coding process and code quality measures to predict project sustainability. 
    
\end{itemize}
\vspace{-0.5mm}

As shown in Table~\ref{baselines}, the final model outperforms all other baselines in terms of AUC and precision.
Although the recall is lower than \textit{Random Forest}, the higher AUC indicates a better overall performance at all thresholds (and thus better fitness to data).
Without the common variables that are studied in other work, the AUC of our model is maintained at 0.71. We compare our findings on the common variables with other studies in Section~\ref{comparefinding}.
Among various dimensions of variables, \textit{Stability of Effort} and \textit{General OSS Experience} are most effective in predicting sustainability.

\begin{table}[h!t]
\footnotesize
\centering
\vspace{-0.25cm}
\caption{AUC/Precision/Recall of our approach and baselines.}
\vspace{-0.3cm}
\label{baselines}
\begin{tabular}{l l l l}
\toprule
Models& AUC/P/R & Models & AUC/P/R \\
\midrule
Logistic Regression & 0.67/0.61/0.39 & Random Forest&0.65/0.54/0.58\\
\midrule
Common Variables&0.66/0.58/0.31& Other Variables&0.71/0.62/0.51\\
\midrule
Cumulative Effort&0.59/0.55/0.25 &Stability of Effort & 0.67/0.58/0.49\\
Level of Concentration & 0.56/0.51/0.15&OSS Experience & 0.62/0.57/0.32\\
Popularity&0.55/0.52/0.17&Opportunity&0.55/0.56/0.09\\
Control Variables&0.61/0.57/0.30&\textbf{Our Model}& \textbf{0.72/0.62/0.52}\\
\bottomrule
\end{tabular}
\vspace{-0.3cm}
\end{table}

As Stănciulescu et al.~\cite{yin2022code} stated in their replication package~\cite{CodeQual60:online}, it takes 5-10 days to collect code-related metrics of their 236 GitHub projects. Since it is impractical to collect code-related metrics for a large number of projects, we compare early participation metrics' and code-related metrics' ability to predict two-year sustainability on their 236 projects. We obtain 0.71 AUC, 0.72 precision, and 0.74 recall with early participation metrics, 0.56 AUC, 0.64 precision, and 0.70 recall with code-related metrics, and 0.68 AUC, 0.71 precision, and 0.77 recall with mixed metrics containing both early participation metrics and code-related metrics, through 10-fold cross-validation. It appears the two-year sustainability of projects cannot be effectively predicted by the code-related metrics.
Adding the code-related metrics to our metrics may lead to the degradation of AUC on the test sets.
Further study is needed to achieve efficient prediction of project sustainability with code-related metrics.

\vspace{-1mm}
\begin{result-rq}{Summary for RQ1:}
Early participation factors from an OSS project can effectively predict long-term sustained activity: our modeling approach for predicting two-year sustained activity can achieve an AUC of up to 0.84 ($m=5, k=6$) on 290,255 GitHub projects.
%Besides, initial contribution pattern is the most crucial indicator of sustained activity.
\end{result-rq}

%In this section, we evaluate the performance of our model in predicting project sustained activity and analyze what kind of factors can facilitate sustained activity of projects. 
%At first, we introduce the selection of studied OSS projects in Sec.~\ref{sec:prosl}. We evaluate the prediction performance under different scenarios in Sec.~\ref{predict}, to demonstrate the effectiveness of our model in prediction.
%Then we study the impact of features on project sustained activity from the three dimensions of willingness, capacity and opportunity in Sec.~\ref{sec:will}, \ref{sec:cap} and \ref{sec:oppor}, respectively.
%We further examine the statistics of features under different project contexts in Sec.~\ref{diffcontext}.
%Finally, we show how to interpret impact of factors for specific projects through case studies in Sec.~\ref{sec:CaseStudy}. 

\subsection{RQ2: Determinants of Sustained Activity}
\label{sec:rq2}
\begin{table*}[htbp]
%\footnotesize
\tablesize
\centering
\tabcolsep=0.16cm
\caption{The definitions of variables and their statistics (median/mean) among different project groups. For each variable, the \textit{negative group} is the group of projects where LIME generates negative local explanations for this variable and the \textit{positive group} is the group of projects where LIME generates positive local explanations for this variable.}
\label{tab:stat1}
\vspace{-3pt}
\begin{threeparttable}
\begin{tabular}{clllrrr}
\toprule
  && Variable&Definition & Negative Group & Positive Group &$|r|$\\
\midrule
\parbox[t]{-2mm}{\multirow{31}{*}{\rotatebox[origin=c]{90}{\textbf{Willingness of Participants}}}}&\parbox[t]{-2mm}{\multirow{9}{*}{\rotatebox[origin=c]{90}{Core}}}
&\#cmt\_c&The average number of commits&14.50/18.17&78.00/108.77&0.77\UpArrowLarge\\
&
&\#pr\_c&The average number of pull requests&0.00/0.00&0.50/1.14&0.34\UpArrowLarge\\
&
&\#issue\_c&The average number of reported issues&4.00/9.04&0.00/0.10&0.81\DownArrowLarge\\
&
&\#iss\_comment\_c&The average number of issue comments&0.00/0.52&10.00/20.73&0.74\UpArrowLarge\\
&
&\#cmt\_comment\_c&The average number of commit comments&1.00/2.39&0.00/0.00&0.46\DownArrowLarge\\
&
&\#iss\_event\_c&The average number of issue events&12.00/33.50&0.00/0.74&0.75\DownArrowLarge\\
&
&\#following\_c&The average number of followed developers&47.00/119.59&2.50/5.05&0.75\DownArrowLarge\\
&
&\#star\_pro\_c&The average number of starred projects&77.00/180.24&4.00/12.99&0.67\DownArrowLarge\\
\cline{3-7}&\parbox[t]{-2mm}{\multirow{9}{*}{\rotatebox[origin=c]{90}{Peripheral}}}
&\#cmt\_p&The average number of commits&0.00/0.34&2.90/4.93&0.64\UpArrowLarge\\
&
&\#pr\_p&The average number of pull requests&0.00/0.00&0.25/0.56&0.24\UpArrowMedium\\
&
&\#issue\_p&The average number of reported issues&1.00/2.94&0.00/0.00&0.62\DownArrowLarge\\
&
&\#iss\_comment\_p&The average number of issue comments&0.00/0.31&0.00/2.11&0.27\UpArrowMedium\\
&
&\#cmt\_comment\_p&The average number of commit comments&0.00/0.11&0.00/0.03&0.04\NoEffect\\
&
&\#iss\_event\_p&The average number of issue events&2.00/7.86&0.00/0.04&0.60\DownArrowLarge\\
&
&\#following\_p&The average number of followed developers&12.33/41.50&0.00/0.11&0.76\DownArrowLarge\\
&
&\#star\_pro\_p&The average number of starred projects&24.00/91.02&0.00/0.66&0.75\DownArrowLarge\\
\cline{3-7}&\parbox[t]{-2mm}{\multirow{7}{*}{\rotatebox[origin=c]{90}{Non-code}}}
&\#issue\_n&The average number of reported issues&4.00/9.29&0.00/0.37&0.75\DownArrowLarge\\
&
&\#iss\_comment\_n&The average number of issue comments&0.00/0.14&1.00/2.38&0.73\UpArrowLarge\\
&
&\#cmt\_comment\_n&The average number of commit comments&0.00/0.10&0.00/0.03&0.11\DownArrowMedium\\
&
&\#iss\_event\_n&The average number of issue events&1.20/4.13&0.00/0.39&0.46\DownArrowLarge\\
&
&\#following\_n&The average number of followed developers&1.00/6.12&12.50/49.63&0.26\UpArrowMedium\\
&
&\#star\_pro\_n&The average number of starred projects&27.90/116.19&1.00/28.90&0.31\DownArrowLarge\\
\cline{3-7}&%\parbox[t]{-2mm}{\multirow{6}{*}{\rotatebox[origin=c]{90}{Effort Stability}}}
&\#cmt\_actday&The number of days with commit activity&6.00/6.00&23.00/27.00&0.87\UpArrowLarge\\
&
&\#cmt\_median&The median number of commits per day&0.00/0.13&0.00/0.22&0.03\NoEffect\\
&
&\#cmt\_front&The number of commits in the first half of observation period&58.00/89.00&11.00/19.00&0.69\DownArrowLarge\\
&
&\#cmt\_end&The number of commits in the second half of observation period&7.00/23.00&9.00/38.00&0.06\NoEffect\\
&
&cmt\_day\_std&The standard deviation of commits per day&2.82/3.75&0.85/0.85&0.87\DownArrowLarge\\
&
&cmt\_dev\_std&The standard deviation of commits per code contributor&0.00/5.20&17.50/28.58&0.53\UpArrowLarge\\
\midrule
\parbox[t]{-2mm}{\multirow{21}{*}{\rotatebox[origin=c]{90}{\textbf{Capacity of Participants}}}}&\parbox[t]{-2mm}{\multirow{7}{*}{\rotatebox[origin=c]{90}{Core}}}
&\#cmt\_all\_c&The average number of commits on all GitHub&162.0/201.0&1541.5/2052.0&0.86\UpArrowLarge\\
&
&\#pr\_all\_c&The average number of pull requests on all GitHub&44.00/96.77&1.00/9.79&0.74\DownArrowLarge\\
&
&\#issue\_all\_c&The average number of reported issues on all GitHub&7.00/17.53&50.00/49.88&0.41\UpArrowLarge\\
&
&\#pro\_c&The average number of owned projects&20.00/19.00&6.50/6.77&0.83\DownArrowLarge\\
&
&\#pro\_oneyear\_c&The average number of one-year sustained owned projects&0.00/0.23&3.00/3.53&0.81\UpArrowLarge\\
&
&\#pro\_twoyear\_c&The average number of owned two-year sustained projects&0.00/0.10&2.00/2.60&0.80\UpArrowLarge\\
&
&\#follower\_c&The average number of followers&6.00/7.35&80.00/185.72&0.87\UpArrowLarge\\
\cline{3-7}&\parbox[t]{-2mm}{\multirow{7}{*}{\rotatebox[origin=c]{90}{Peripheral}}}
&\#cmt\_all\_p&The average number of commits on all GitHub&0.0/17.8&876.5/1462.2&0.75\UpArrowLarge\\
&
&\#pr\_all\_p&The average number of pull requests on all GitHub&23.00/75.32&0.00/0.09&0.83\DownArrowLarge\\
&
&\#issue\_all\_p&The average number of reported issues on all GitHub&0.00/0.42&21.60/33.66&0.76\UpArrowLarge\\
&
&\#pro\_p&The average number of owned projects&15.00/13.99&0.00/0.55&0.78\DownArrowLarge\\
&
&\#pro\_oneyear\_p&The average number of owned one-year sustained projects&0.00/0.00&1.50/2.17&0.73\UpArrowLarge\\
&
&\#pro\_twoyear\_p&The average number of owned two-year sustained projects&0.00/0.01&1.00/1.38&0.57\UpArrowLarge\\
&
&\#follower\_p&The average number of followers&0.00/6.68&13.00/60.55&0.64\UpArrowLarge\\
\cline{3-7}&\parbox[t]{-2mm}{\multirow{7}{*}{\rotatebox[origin=c]{90}{Non-code}}}
&\#cmt\_all\_n&The average number of commits on all GitHub&0.0/200.6&799.7/1230.7&0.55\UpArrowLarge\\
&
&\#pr\_all\_n&The average number of pull requests on all GitHub&42.00/88.58&0.00/1.27&0.84\DownArrowLarge\\
&
&\#issue\_all\_n&The average number of reported issues on all GitHub&0.00/5.58&55.00/54.98&0.77\UpArrowLarge\\
&
&\#pro\_n&The average number of owned projects&20.00/17.57&0.00/3.76&0.78\DownArrowLarge\\
&
&\#pro\_oneyear\_n&The average number of owned one-year sustained projects&0.00/0.16&1.00/2.05&0.67\UpArrowLarge\\
&
&\#pro\_twoyear\_n&The average number of owned two-year sustained projects&0.00/0.86&0.00/0.30&0.25\DownArrowMedium\\
&
&\#follower\_n&The average number of followers&0.00/5.36&55.20/123.72&0.74\UpArrowLarge\\
\midrule
\parbox[t]{-2mm}{\multirow{5}{*}{\rotatebox[origin=c]{90}{\textbf{Oppo.}}}}
&
&\#iss\_open&The number of open issues three months after project creation&3.00/7.00&0.00/2.00&0.42\DownArrowLarge\\
&
&iss\_open\_ratio&The ratio of open issues three months after project creation&0.00/0.04&0.63/0.64&0.82\UpArrowLarge\\
&
&\#GFI&The number of \emph{good first issues} three months after project creation&0.00/2.00&0.00/0.00&0.18\DownArrowMedium\\
&
&\#line\_readme&The number of README.md lines&44.00/55.00&23.00/87.00&0.03\NoEffect\\
&
&\#line\_contributing&The number of CONTRIBUTING.md lines&0.00/0.00&0.00/40.00&0.23\UpArrowMedium\\
\midrule
\parbox[t]{-2mm}{\multirow{10}{*}{\rotatebox[origin=c]{90}{\textbf{Control Variables}}}}&
&show\_comp\_c&The ratio of core developers showing affiliated companies/institutions&0.00/0.12&1.00/0.86&0.74\UpArrowLarge\\
&
&\#org\_c&The average number of GitHub organizations core developers belong to&1.50/2.30&0.00/0.32&0.52\DownArrowLarge\\
\cline{3-7}&
&show\_comp\_p&The ratio of peripheral developers showing affiliated companies/institutions&0.00/0.00&0.75/0.71&0.74\UpArrowLarge\\
&
&\#org\_p&The average number of GitHub organizations peripheral developers belong to&0.00/0.07&0.50/1.22&0.46\UpArrowLarge\\
\cline{3-7}&
&show\_comp\_n&The ratio of non-code contributors showing affiliated companies/institutions&0.00/0.08&0.83/0.69&0.70\UpArrowLarge\\
&
&\#org\_n&The average number of GitHub organizations non-code contributors belong to&0.30/1.27&0.00/0.40&0.29\DownArrowMedium\\
\cline{3-7}&
&type&The type of project owner account (0: organization, 1: user)&1.00/1.00&0.00/0.00&0.87\DownArrowLarge\\
&
&\#star&The number of stars three months after project creation&5.00/58.00&0.00/0.00&0.77\DownArrowLarge\\
&
&\#fork&The number of forks three months after project creation&1.00/1.00&0.00/9.00&0.12\DownArrowMedium\\
&
&\#member&The number of project members three months after project creation&0.00/0.00&1.00/3.00&0.58\UpArrowLarge\\

\bottomrule
\end{tabular}
\begin{tablenotes}
\footnotesize
\item  $r$ represents the effect size $(r=z/\sqrt{N})$~\cite{fritz2012effect}; \UpArrow, \UpArrowMedium, \UpArrowLarge~means the positive group has higher median for this variable and the effect size is small, medium, and large, respectively (\DownArrow, \DownArrowMedium, \DownArrowLarge~vice versa).
\end{tablenotes}
\end{threeparttable}
\vspace{-8pt}
\end{table*}
%In this Section, we report the results obtained from applying the methodology in Sec.~\ref{lime} to the model trained with $m = 3$, $t = 2$ and $k = 1$.
%In this setting, the model uses projects' data in the first three months to predict whether a GitHub project will have two-year sustained activity (i.e., a time span of at least two years between the first and last commits, and $\geqslant$12 months with commit activities within the first two years).
%The choices of $t$ and $k$ are based on the intuition that two years are a sufficiently long time period for ``sustained development'' and commit activities in half of the 24 months is a minimum bar; the results are also stable across different $t$s and $k$s (see replication package, Sec.~\ref{sec:data}).

Table~\ref{tab:stat1} summarizes the results we obtained from applying LIME to the trained model using all projects in our dataset.
We use Mann–Whitney U test~\cite{mann1947test} to test all the 64 variables in both groups and for all variables the $p$-value is less than $0.00078$ ($<0.05/64$, with Bonferroni correction~\cite{dunn1961multiple}).
More than half of the tests have an effect size of more than 0.5, which indicates a large difference in the variable values between the negative and positive groups.
In this Section, we discuss the effect of variables in detail.

\textit{\textbf{Willingness of Participants.}}
Similar to Section~\ref{sec:feature}, we discuss the effect of variables in the three sub-dimensions.

\vspace{-1mm}
\begin{itemize}[leftmargin=20pt]
\item[\scriptsize{WP}\small{1}]
\textbf{Cumulative Effort Amount:} 
As expected, we find that the willingness of participants in code contributions (\Code{\#cmt\_\{c|p\}}, \Code{\#pr\_\{c|p\}}) is a positive factor for long-term sustained activity in project early stages. 
The effect of participant willingness in submitting issues (\Code{\#issue\_\{c|p|n\}}) and adding comments to commits (\Code{\#cmt\_comment\_\{c|p|n\}}) is generally negative. 
Our interpretation is that too many early issues and too much interaction effort in code contributions may diverge participants' attention, harming sustainability in the long term.
However, issue discussions (\Code{\#iss\_comment\_\{c|p|n\}}) carry long-term benefits. 
The reason could be that discussions usually clarify the problem and solution before issue resolvers contribute, which thus reduces unnecessary effort. 
\item[\scriptsize{WP}\small{2}]
\textbf{Stability of Effort:}
We discover that a steady stream of commits (\Code{\#cmt\_actday}) has a large positive effect on project long-term sustained activity. Furthermore, we find that making most of the commits in the first half of the observation period (\Code{\#cmt\_front}) and having unevenly distributed commits over time (\Code{cmt\_day\_std}) has a large negative effect.
%This is expected because early investment indicates intrinsic motivations to participate, which drives them to sustain contributions in the future~\cite{DBLP:journals/mansci/RobertsHS06}.
However, the uneven distribution of workload among developers (\Code{cmt\_dev\_std}) carries a large positive effect instead, which may be because it indicates the presence of highly competent developers and several peripheral developers, who are often vital for the incubation of successful OSS.
\item[\scriptsize{WP}\small{3}]
\textbf{Level of Concentration:}
We discover that, for core developers and peripheral developers, both their average number of followed developers (\Code{\#following\_\{c|p\}}) and their average number of starred projects (\Code{\#star\_pro\_\{c|p\}}) have a negative effect on long-term sustained activity.
However, for non-code contributors, their average number of followed developers (\Code{\#following\_n}) has a positive effect.
This indicates that the divergent attention of core and peripheral developers may limit their effort and willingness in one single project, harming its sustainability.
However, participation (discussions in issues, \Code{iss\_comment\_n}) of such non-code contributors instead carries a positive effect, maybe because those with a broad interest can provide interesting and helpful suggestions.
\end{itemize}
\vspace{-1mm}

To summarize, early participation from non-code contributors and a large and steady coding effort has positive effects on project sustainability, while too many early issues, commit discussions, and divergent attention of code contributors can carry negative effects.

\textit{\textbf{Capacity of Participants.}}
We report the effect of variables in the two sub-dimensions as in Section~\ref{sec:feature}.

\vspace{-1mm}
\begin{itemize}[leftmargin=20pt]
\item[\scriptsize{CP}\small{1}]
\textbf{General OSS Experience:}
We find that the number of previous commits and issues (\Code{\#cmt\_all\_\{c|p|n\}}, \Code{\#issue\_all\_\{c| p|n\}}) on GitHub generally has a positive effect on long-term sustainability, as it measures prior OSS development experiences.
However, the number of previous PRs on GitHub (\Code{\#pr\_all\_\{c|p|n\}}) has a negative effect.
This may be because PRs are often submitted to external projects and thus, indicate the developers' capacity is distributed among many projects. 
Furthermore, we find that the number of owned projects (\Code{\#pro\_\{c|p|n\}}) has a large negative effect for all three types of participants, while the number of owned projects with one-year or two-year sustained activity (\Code{\#pro\_oneyear\{c|p\}}, \Code{\#pro\_twoyear\{c|p\}}) has a large positive effect for code contributors.
This indicates that prior experience in incubating long-term sustainable projects is a key indicator of code contributors' capacity in sustaining an OSS project, but their capacity does not help if distributed among too many projects.
Also, owning too many OSS projects may imply one's haphazardness in creating projects that are often abandoned rapidly.
\item[\scriptsize{CP}\small{2}]
\textbf{Popularity:}
As expected, the number of followers (\Code{\#follower\_ \{c|p|n\}}) has a large positive effect on long-term sustained activity for all three types of participants, as a developer's popularity among other GitHub developers is a good indicator of their ability to develop useful or interesting projects.
\end{itemize}
\vspace{-1mm}

To summarize, we show that participants' prior OSS experience in coding and incubating long-term sustainable projects, and their popularity among other developers, are positive indicators of their capacity in fostering long-term sustainability.

\textit{\textbf{Opportunity of Contribution.}}
The opportunity of contribution, as measured by the ratio of open issues (\Code{iss\_open\_ratio}) and the number of lines of CONTRIBUTING file (\Code{\#line\_contributing}), has a positive effect on long-term sustained activity.
One possible explanation is that projects with higher open issue ratios have a better chance of attracting new contributions. Besides, more detailed contribution guidelines can reduce barriers for newcomers to contribute.
However, more open issues and GFIs (\Code{\#iss\_open}, \Code{\#GFI}) are usually accompanied by more issues, both of which have a negative effect on sustainability activities.

\textit{\textbf{Control Variables.}}
Consistent with previous findings~\cite{DBLP:conf/sigsoft/ValievVH18,kula2019life}, we find that organizational projects are more likely to have long-term sustained activity.
Among other control variables, the number of project members (\Code{\#member}) and the ratio of participants showing their affiliated companies/institutions on their overview page (\Code{show\_comp\_\{c|p|n\}}) have a positive effect, while the number of stars (\Code{\#star}) and the number of forks (\Code{\#fork}) have a negative effect; the number of organizations (\Code{\#org\_\{c|p|n\}}) has a mixed effect for participants.
Our interpretation is that the positive factors often signal a larger development team (e.g., hosted under an organization, members, affiliations), so they are more likely to sustain.
However, early stars, forks, and the number of GitHub organizations do not serve as an indicator of initial projects' future sustained activity. 

\begin{figure}[t]
\vspace{-0.2cm}
\centering
%\captionsetup[subfloat]{labelsep=none,format=plain,labelformat=empty} 
\subfloat
{
\centering
\includegraphics[scale=0.20]{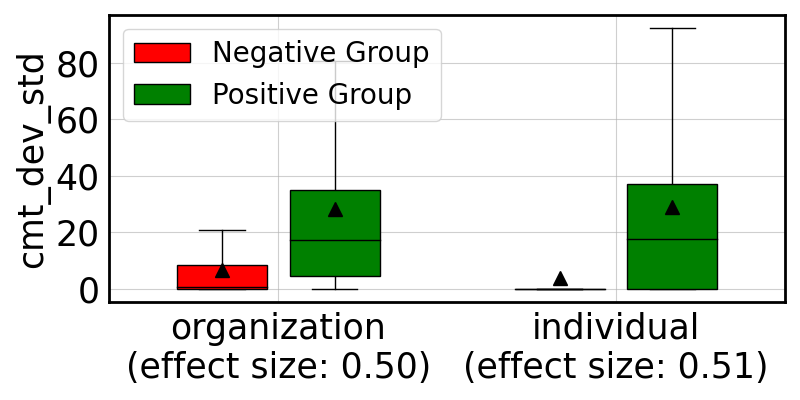}
}
\subfloat
{
\centering
\includegraphics[scale=0.20]{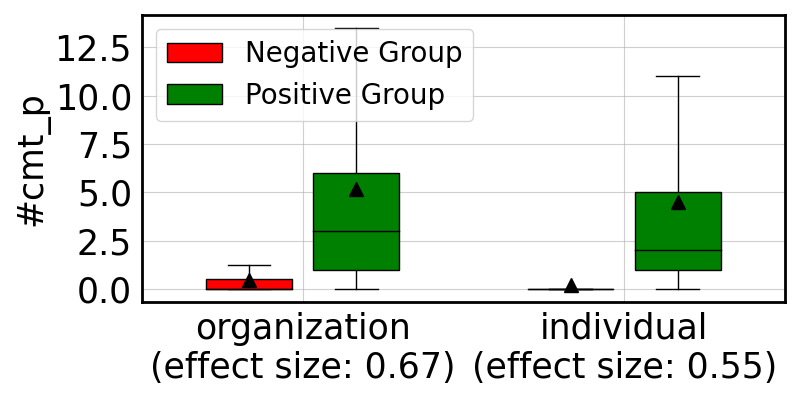}
}

\subfloat
{
\centering
\includegraphics[scale=0.20]{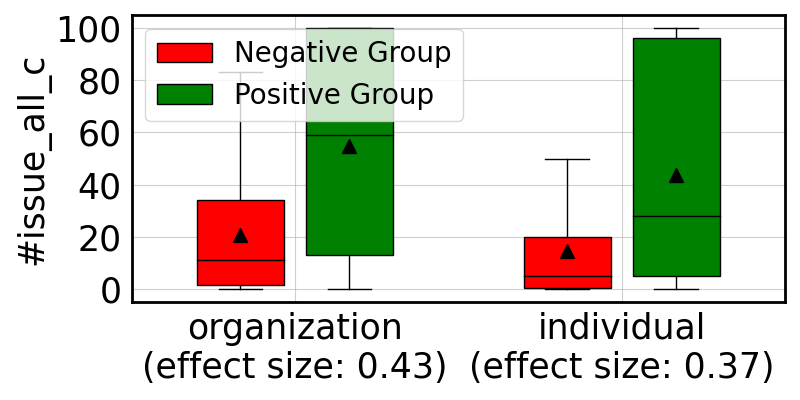}
}
\subfloat
{
\centering
\includegraphics[scale=0.20]{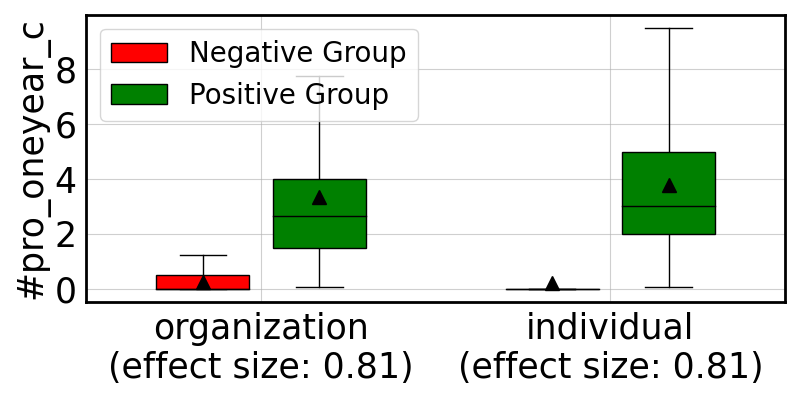}
}
\vspace{-0.3cm}
\caption{The distribution of variables within the negative (red) and positive (green) groups for projects owned by organization accounts and individual accounts, respectively.}
\label{vardistibution}
\vspace{-0.4cm}
\end{figure}

As the largest effect size (0.87) comes from the \Code{type} variable, we are interested in whether the effects of variables differ between projects owned by organizational and individual accounts.
For each variable, we divide projects by \Code{type} and separately investigate the differences between the positive group and the negative group.
After an investigation, we find that the most significant differences come from the variables in \textit{Stability of Effort} (i.e., \Code{cmt\_dev\_std}), \textit{Cumulative Effort Amount} (i.e., \Code{\#cmt\_p}), and \textit{General OSS Experience} (i.e., \Code{\#issue\_all\_c} and \Code{\#pro\_oneyear\_c}), as shown in Figure~\ref{vardistibution}. Compared with individual projects, higher \Code{cmt\_dev\_std} is required to have a positive effect on the sustained activity of organizational projects.
This indicates that sustainable organizational projects generally require not only concentrated contributions from core developers, but also commitments from peripheral developers.
%concentrated contributions from core developers (i.e., a high level of commitment) while sustainable individual projects need not be so centralized. 
Besides, organizational projects have a larger \Code{\#cmt\_p} in the group where it has a positive effect, indicating that organizational projects may require more commits from peripheral developers to demonstrate their likelihood of being sustainable.
On the other hand, sustainable organizational projects generally require more OSS experience (i.e., higher \Code{\#issue\_all\_c} and \Code{\#pro\_oneyear\_c}) from core developers. 

\vspace{-0.1cm}
\begin{result-rq}{Summary for RQ2:}
We find that the following early participation factors promote long-term project sustained activity:
\begin{itemize}[leftmargin=10pt]
    \item \textbf{Willingness:} Steady attention and commitment from developers; participation from non-code contributors.
    \item \textbf{Capacity:} Rich OSS experience, experience in sustaining OSS projects, and many followers of the code contributors.
    \item \textbf{Opportunity:} A high ratio of open issues; detailed contribution documentation.
\end{itemize}
To be sustainable, organizational projects require more experienced core developers and more active peripheral developers compared to individual projects.

\end{result-rq}
\vspace{-0.1cm}

\vspace{-2mm}
\section{Discussion}

\vspace{-0.5mm}
\subsection{Comparing Findings with Prior Work}\label{comparefinding}
\vspace{-0.5mm}

As already mentioned in Section~\ref{sec:related-work}, our work takes a unique perspective, which is different from all prior work to the best of our knowledge.
However, due to similarities in operationalization, our study has yielded some overlapping findings with prior work.

In terms of participant willingness, we identify the willingness in code contributions (measured by the average number of commits in each participant group) as a positive factor.
This finding extends the results of Yin et al.~\cite{DBLP:conf/sigsoft/YinCXF21} and Stănciulescu et al.~\cite{yin2022code} showing the total effort of development (measured by the number of commits) as a positive factor.
However, operationalizing development effort with the number of commits could be problematic in mature projects in which merging large contributions into a single commit is a better practice, turning the number of commits as a negative factor~\cite{DBLP:conf/sigsoft/ValievVH18}.
The willingness in opening issues is identified as a negative factor, which is in line with Valiev et al.~\cite{DBLP:conf/sigsoft/ValievVH18} reporting the number of early issues as a risk factor of Python package activity.
Consistent with the observation of Stănciulescu et al.~\cite{yin2022code}, our results also show that a steady stream of commits (\Code{\#cmt\_actday}) has a large positive effect on project long-term sustained activity.
For the remaining participant factors, we are unaware of any prior empirical investigation in the context of OSS project sustainability (e.g., the level of concentration and experiences of prior OSS incubation).

For the control variables also investigated in prior work (Table~\ref{relcom}), their effect is also similar, including the presence of an organization~\cite{DBLP:conf/sigsoft/ValievVH18}, and development team size~\cite{DBLP:journals/jais/Chengalur-SmithSD10,DBLP:conf/sigsoft/ValievVH18,DBLP:conf/sigsoft/YinCXF21,wang2012survival,DBLP:journals/jss/MidhaP12,DBLP:journals/infsof/SamoladasAS10}.
%both of which exert a positive effect at project early stage.
We believe the confirmation of prior findings from our control variables could be a signal of validity for our methodology.

%However, in later stages, more developers may increase effort and emphasis needed to manage and coordinate the team, which may ultimately lead to a decrease in project activity~\cite{DBLP:journals/jss/MidhaP12}.

\vspace{-1mm}
\subsection{Practical Implications}\label{secimplication}

%In this Section, we discuss, from the lens of sustainability, how the results from our paper support certain practices and strategies in OSS project incubation, how to identify promising OSS projects in their early stages, and what implications  our study has for future research.

\subsubsection{Incubating Sustainable OSS Projects}

It is hard to kickstart an OSS project; it is even harder to incubate an OSS project with long-term sustainability.
Despite the presence of many guidelines in the grey literature~\cite{OSS-Start1, OSS-Start2, OSS-Start3, OSS-Start4}, they often do not explicitly mention the sustainability problem, nor do they provide empirical evidence supporting the effectiveness of their suggestions on sustainability.
Based on our findings, we discuss some practices that may help the incubation of sustainable OSS projects.

\textbf{Stability of Contributions.} 
Our results for Willingness show that not only the large number of commits (\Code{\#cmt\_\{c|p\}}, \Code{\#pr\_\{c|p\}}, \Code{\#cmt\_actday}) studied before is important for project sustainability~\cite{DBLP:conf/sigsoft/YinCXF21,yin2022code}, but also continued investment (\Code{\#cmt\_day\_std}) and concentration (small \Code{\#following\_\{c|p\}} and \Code{\#star\_pro\_\{c|p\}}) of code contributors. 
To maintain stable contributions, we suggest that initial project maintainers: 1) set clear goals for the project so that developers clearly know what code contributions they can make, thus the excessive time and energy spent on the debate after the code is submitted (\Code{\#cmt\_comment\_\{c|p|n\}}) can be reduced; 2) provide positive feedback (e.g., timely and friendly issue responses, \Code{\#iss\_comment\_\{c|p|n\}}) to other developers to motivate them to continue contributing; and 3) watch for signs of developer attrition in advance: a steady decline in contributions (\Code{\#cmt\_day\_std}) and a shift of interest to other projects (\Code{\#pro\_\{c|p|n\}}).

\textbf{Welcome to Newcomers.}
Previous studies~\cite{2012ICSE-Zhou-What,2020FSE-Tan-First} point out that reporting and resolving issues (\Code{iss\_open\_ratio}) is an effective way for newcomers to join projects.
We further find that issue comments from non-code contributors (\Code{\#iss\_comment\_n}) are positively related to sustained activity, even if they do not resolve the issue. It may suggest that commenting on issues is also an appropriate way for external contributors to interact with the OSS community. Such interactions can provide feedback to development or lead to future contributions, and thus should not be ignored by project maintainers.
Some projects may overlook the role of contribution documentation in the early stage~\cite{Contribu75:online,AddedCON9:online}. 
Newcomers may find it difficult to contribute, and even wonder if the project is actively accepting new PRs~\cite{Contribu75:online}.
Our results suggest that maintainers should provide detailed contribution documentation (\Code{\#line\_contributing}) early on, which not only lowers the barriers for newcomers to contribute, but also expresses the attitude of welcoming newcomers.

\textbf{Competent Team.}
Different developers have different roles in a project, and the required effort and experience may vary accordingly. As far as we know, we are the first to characterize the requirements of sustainable projects for developers in different roles.
Our results suggest that projects, especially organizational projects, should pay attention to the retention of experienced core developers (with higher \Code{\#issue\_all\_c}, \Code{\#pro\_oneyear\_c}) and active peripheral contributors (for higher \Code{cmt\_dev\_std}, \Code{\#cmt\_p}).

\textbf{Activity Management.} Previous studies find the positive effect of making commits and the negative effect of reporting issues on project sustainability~\cite{DBLP:conf/sigsoft/YinCXF21,yin2022code,DBLP:conf/sigsoft/ValievVH18}.
Our results further demonstrate that direct code contributions (\Code{\#cmt\_\{c|p\}}, \Code{\#pr\_\{c|p\}}) have a positive effect on sustained activity, while other types of activities (\Code{\#issue\_\{c|p|n\}}, \Code{\#cmt\_comment\_\{c|p|n\}}, \Code{\#iss\_event\_\{c|p|n\}}, \Code{\#} \Code{GFI}) do not. 
Developers can engage in a variety of activities on GitHub projects, however, their energy is limited.
Therefore, we suggest that initial project maintainers should be careful not to let these ``process'' activities take up too much time and effort, since some non-code activities take resources and sometimes slow down development activities~\cite{DBLP:conf/sigsoft/ValievVH18}. In the initial phase of a project, it may be more appropriate to focus on ``getting things done'' (e.g., coding and documenting, as measured by \Code{\#cmt\_\{c|p\}}, \Code{\#pr\_\{c|p\}}, \Code{\#line\_contributing}), than to focus on ``process'' (reflected by \Code{\#issu} \Code{e\_\{c|p|n\}}, \Code{\#cmt\_comment\_\{c|p|n\}}).
The latter may introduce unnecessary overhead and harms the motivation of the early development team. Some guidelines suggest labeling \emph{good first issues} as soon as the project launches~\cite{OSS-Start4}. However, only 2.7\% of the projects in our dataset adopted the GFI mechanism in their first three months. Even in the projects that have adopted it, it may not be as helpful in attracting high-capacity contributors (\Code{\#GFI} has a small negative effect on sustained activity). Further efforts are needed to make it work to help initial projects actually attract valuable newcomers.

\vspace{-2mm}
\subsubsection{Identifying Potential Sustainable Projects}
As open source grows, it is likely that multiple GitHub projects belong to the same subfield or provide similar functionality. For \textit{newcomers} aspiring to open source development, taking part in a project that will die on the vine is not only frustrating, but also a waste of time and energy. For \textit{developers} deciding whether to use a third-party library, \textit{end users} deciding whether to use an 
OSS application, and \textit{companies/foundations} deciding whether to invest in an initial project, it is also important to identify potential long-term sustainable projects from a large number of candidates.

\textbf{Indicators}.
Based on our results, the following indicators can be used to recognize potential sustainable projects at their early stage:
1) the contributors make large (check \Code{\#cmt\_\{c|p\}}, \Code{\#pr\_\{c|p\}}) and steady (check \Code{\#cmt\_actday}, \Code{\#cmt\_day\_std}, \Code{\#cmt\_front}) code contributions; 2) the code contributors have focused attention (check
\Code{\#following\_\{c|p\}}, \Code{\#star\_pro\_\{c|p\}}); 3) the participants have rich experience in making commits, reporting issues, and incubating long-term sustainable projects (check \Code{\#cmt\_all\_\{c|p|n\}}, \Code{\#issue\_all\_\{c|p|n\}}, \Code{\#pro\_oneyear\_\{c|p|n\}}, \Code{\#pro\_twoyear\_\{c|p}
\Code{\}}); 4) the participants have many followers (\Code{\#follower\_\{c|p|n\}}); 5) the project provide a high ratio of open issues (\Code{iss\_open\_ratio}) and detailed contribution documentation (check \Code{\#line\_contributi} \Code{ng}); and 6) the project is owned by an organization account (check \Code{type}) and has many members (\Code{\#member}). Among the signals, only \Code{\#cmt\_\{c|p\}}, \Code{\#cmt\_actday}, \Code{type} and \Code{\#member} are presented by other work. Our results also reveal that signals in different contexts may need to be adjusted, e.g. sustainable organizational projects are more demanding for developers.

\vspace{-0.2cm}
\subsubsection{Future Research Directions}
\,\\
\textbf{Modeling Tools}.
Our study suggests the development of tools for identifying potentially sustainable OSS projects and recommending interventions for improving the long-term sustainability in the project's early stages.
Our methodology and results provide a promising starting point and interesting implications for tool design.
For example, the methodology can be implemented as a tool to aid judgment. 
The XGBoost model can predict the future sustainable probability of an initial project, and the interpretable LIME model can further provide the contribution value of each variable to the sustainability of the project.
Newcomers can use the XGBoost prediction model to quickly identify potentially sustainable projects from the flood of GitHub projects, while project maintainers can monitor their projects and learn how to derive intervention on the projects from the LIME model's interpretation.

\textbf{Visualization Tools}.
Also, the determinants we identified are often not obvious from the common sources of information, while those common ones (e.g., the number of stars) are not good indicators of long-term sustained activity.
Therefore, it should also be helpful to have a tool that aggregates and visualizes important sustainability determinants into a central information panel.
%However, future work is needed to overcome the implementation barriers posed by GitHub and systematically evaluate the tool's usefulness.

\textbf{Other Metrics}.
Apart from the early participation factors that we have investigated, factors related to software artifact quality (e.g., code, tests) may also have an important effect on long-term sustainability.
As well-known, high-quality code and tests facilitate long-term maintainability, so having them is generally recommended for OSS project launching~\cite{OSS-Start1,OSS-Start2}.
Previous studies have established correlations between code quality metrics and OSS sustainability~\cite{yin2022code, DBLP:journals/is/Ghapanchi15}. 
But how to apply these code-related metrics to learning and predicting sustainability for a larger number of projects, and whether there are other code metrics that can predict project sustainability more efficiently, deserve future investigations.
In addition, projects of different sizes, complexities, and domains may emphasize different best practices to achieve long-term sustainability. We leave the exploration of this aspect to future research.

\textbf{Organizations and Companies}.
Finally, since the presence of organizations and companies has largely positive effects in our analysis, we are interested in the underlying mechanisms and the roles that organizations and companies may play during OSS launching.
This may provide explanations on why early participation from organizations and companies helps long-term sustained activity.

\vspace{-1mm}
\subsection{Limitations}
\label{sec:limitations}

\textbf{Correlation and Causation.}
We applied LIME to explain the contribution of variables to the predicted probability from a non-linear ML model, where LIME locally fits a linear regression model for each sample. 
Similar to regression modeling approaches, this approach only detects correlations and does not guarantee causality~\cite{sparks2010regression}. The correlations we discover in this study can serve as hypotheses for future studies that aim to establish causality relationships.

\textbf{Construct Validity.} 
First, our definition of sustained activity may not cover all cases in which developers consider a project ``sustainable''. 
%For example, projects may become inactive because they are ``completed'' with no new requirements~\cite{DBLP:conf/sigsoft/CoelhoV17} or ``revive'' after a long period of inactivity~\cite{DBLP:conf/esem/AvelinoCVS19}.
To mitigate this risk, we ensure that the determinants identified from our methodology are stable across different sustained activity definitions (i.e., $t$, $m$ and $k$ choices).
Second, while most of the determinants remain stable for one- to four-year sustained activity (see the results in the replication package), we acknowledge that several determinants for longer-term sustained activity may differ, and projects that aim to achieve sustainability over extended periods may need to adjust some best practices.
Third, our dataset consists of projects with a certain level of activeness and popularity, which may limit the generalizability of our findings to projects that do not exhibit clear signals of long-term development interest. However, it is worth noting that over half (58\%) of the projects in our study failed to achieve sustainability beyond two years (see Table~\ref{tab:age}). The focus of our research is precisely on these projects that demonstrate a willingness for long-term development but face challenges. This mitigates the risk of our results being inapplicable to the target projects.
Fourth, our analysis solely focuses on variables available on GitHub, which may not comprehensively capture developers' willingness, capacity, and opportunity. %For instance, online discussions or promotional activities related to a project can serve as indicators of developers' willingness to participate in the project. Prior research indicates a relationship between promotion on social media websites and the popularity of GitHub projects~\cite{DBLP:journals/jss/BorgesV18}. The investigation of the impact of social media website activities on project sustainability remains an area for future exploration.
Fifth, our dataset selection approach may still involve projects with no intention of long-term sustainability, and we mitigate this risk with aggressive filtering criteria and a manual evaluation.
The final threat comes from the reliability of GHTorrent, which cannot be considered a full replica of GitHub according to previous analysis~\cite{gousios2013ghtorent}.
However, for our study protocol, we believe it is still the best choice and it would be difficult to sample and collect data using only the GitHub API. 

\textbf{External Validity.} 
Our study only focuses on GitHub projects and does not consider fork projects. The findings may not generalize to OSS projects hosted elsewhere or OSS projects that are initiated as a fork of other projects. 
To mitigate the generalizability threat to other OSS projects, we carefully selected a large-scale dataset of top-ranked GitHub projects covering a diverse range of software domains.
Thus, we believe the findings and implications should be applicable to a broad audience of OSS developers.
The generalizability threat to fork projects falls out of the scope of our study and future work is needed to investigate sustainable ``fork''.
Furthermore, the results and recommendations presented in this paper are based on a general analysis of variable contributions, and can be applied to most projects. 
However, they may vary slightly for projects with different backgrounds.
%For example, as discussed in Section~\ref{sec:rq2}, different owner account types may necessitate different best practices. 
We suggest that initial project maintainers seeking project-specific practical suggestions follow a similar methodology by utilizing our interpretive LIME model, which leads to more precise and tailored recommendations.

\section{Conclusion}
\label{sec:conc}

In this paper, we propose an XGBoost-based model that uses early participation factors to predict project's long-term sustained activity %and present an empirical study that quantitatively investigates the early participation determinants of long-term sustained activity 
on a dataset of 290,255 GitHub projects.
More specifically, we measure the willingness, capacity, and opportunity of participants in the first three months of projects, train an XGBoost model to predict projects' two-year sustained activity, and interpret the contribution of variables to the prediction outcome using LIME.
We confirm that early participation factors can indeed be used for prediction and a number of determinants are identified from our studied factors.
Our study provides quantitative evidences supporting the effectiveness of certain practices and strategies on long-term OSS sustainability, which benefits OSS practitioners. We also provide insights for identifying potential long-term sustainable projects.
However, the topic of fostering OSS sustainability early is very broad: other potential factors and their interplays may be missed by our study and future research is needed for their investigation.

\vspace{-1mm}
\section{Data Availability}
\label{sec:data}

A replication package is at:
\url{https://zenodo.org/record/8098994}

\section*{Acknowledgment}
This work is supported by the National Natural Science Foundation of China Grant 61825201, 62332001, and 62141209.

\bibliographystyle{ACM-Reference-Format}
\bibliography{references}

\end{document}